\documentclass[citeautoscript,prx,aps,twocolumn,showpacs,showkeys,superscriptaddress,amsmath,amssymb]{revtex4} 

\usepackage{graphicx}
\usepackage{psfrag}
\usepackage{epsfig}
\usepackage{color}
\usepackage{subfigure}
\definecolor{dred}{rgb}{0.7,0.0,0.0}
\usepackage{dcolumn}
\usepackage{bm}

\begin{document}

%
%

\title{Minimal-size Real Space $d$-wave Pairing Operator in CuO$_2$ Planes}

\author{Adriana Moreo}
\author{Elbio Dagotto}
 
\affiliation{Department of Physics and Astronomy,University of Tennessee,
Knoxville, TN 37966, USA} 
\affiliation{Materials Science and Technology Division,
Oak Ridge National Laboratory,Oak Ridge, TN 37831, USA}

\date{\today}

\begin{abstract}
{A novel minimal-size pairing operator $\Delta^{\dagger}_{D0}$ with $d$-wave 
symmetry in CuO$_2$ planes is introduced. 
This pairing operator creates on-site Cooper pairs at the four oxygens that surround a copper atom. 
Via the time evolution of $\Delta^{\dagger}_{D0}$, an additional inter-orbital pairing operator $\Delta^{\dagger}_{Dpd}$ with $d$-wave symmetry 
is generated that pairs fermions located in a Cu and its four surrounding O's. The subsequent time evolution of
$\Delta^{\dagger}_{Dpd}$ generates an intra-orbital $d$-wave pairing operator $\Delta^{\dagger}_{Dpp}$ involving 
the four O atoms that surround a Cu, as well as the $d$-wave operator $\Delta^{\dagger}_D$ 
traditionally used in single-band models for cuprates.   
Because we recover the larger size operators extensively used in the three-orbital Hubbard model, we suggest that long-range order using the canonical extended 
operators occurs together with long-range order in the new minimal operators. However, our minimal $d$-wave operators could be more practical to study $d$-wave superconductivity because in the finite-size relatively small systems accessible to computational techniques
it is easier to observe long-range order using local operators. Moreover, 
an effective model with the usual tight-binding hopping of the CuO$_2$ planes supplemented by an attractive potential $V$  in the $d$-wave channel is introduced. 
Using mean-field techniques 
we show that a paired ground state is stabilized for any finite value of $V$. We observed that 
the values of $V$ that lead to gap sizes similar 
to those in the cuprates are smaller for $d$-wave pairing operators that include 
Cu $d$-orbitals than those that only include $p$-orbitals. 
In all cases the gap that opens in the spectrum has standard $d$-wave 
symmetry. Finally, a simpler effective model is introduced to study the phenomenology 
of multi-orbital $d$-wave superconductors, similarly as the negative-$U$ Hubbard model is 
used for properties of $s$-wave superconductors.}
\end{abstract}
 
 
\maketitle

\section{Introduction} 

The discovery of $d$-wave superconductivity in the high critical temperature cuprates~\cite{dwave1,dwave2} 
started efforts to develop effective Hamiltonians that would allow to study $d$-wave pairing 
in the same way as the negative-$U$ Hubbard model
allows the study of $s$-wave pairing in standard BCS superconductors~\cite{uneg1,uneg2,uneg3}. 
Previous efforts focused on single-orbital systems with on-site Coulomb repulsion together with an effective 
attractive nearest-neighbors potential~\cite{dw1,dw2}, hardcore dimers~\cite{dw3}, 
or via the  phenomenological addition of a term proportional to the square of the
nearest-neighbor hoppings~\cite{dw4}. These models were difficult to study, parameters needed 
to be fine tuned, and actual numerical evidence of long-range $d$-wave pairing
correlations has been elusive~\cite{dw3bis}. The contribution of orbital degrees of freedom to the symmetry 
of the pairing operator came to the foreground when superconductivity 
was observed in iron-based pnictides and selenides~\cite{fe1,fe2,fe3,fe4} and recently 
an effective model with on-site inter-orbital attraction was presented~\cite{dw5}. While
relatively easy to study, inter-orbital same-site pairing operators are considered to be less likely 
to develop long-range order than their intra-orbital counterparts involving the same orbital but
at  different sites. For all these reasons, it is still important to find alternative and practical
intra-orbital pairing operators with $d$-wave symmetry. 

In addition, recent angle-resolved photoemission experiments
using Bi$_2$Sr$_2$CaCu$_2$O$_{8+\delta}$ indicated a novel ``starfish'' shape of the superconducting 
pairs with a very short length -- of the order of one lattice space -- in the antinodal 
direction~\cite{smallpairs}. This unexpected result appears to be doping independent and 
it may offer clues on the local structure of $d$-wave pairs in the strong coupling regime.
For us these experiments provide additional motivation to reconsider the local form of 
the $d$-wave pairing operators in the cuprates.

Most previous attempts to construct same-orbital effective $d$-wave models, analogous
to the $U<0$ Hubbard model for $s$-wave, relied on single-orbital systems with electrons placed
on sites of a square lattice that mimic only the coppers. 
In the present publication, we aim to explore whether effective models for $d$-wave superconductivity
can be constructed using, instead, the oxygen locations in the more realistic CuO$_2$ lattice. 
It is well known that
holes tend to reside on oxygens due to the charge transfer nature of the cuprates. However, the
vast majority of theory efforts in this context rely on one-orbital Cu-only models, such as the $t-J$ and one-orbital 
Hubbard. Only recently computational efforts are studying
the full CuO$_2$ models, with both Cu and O
incorporated, and interesting results such as stripes have already been unveiled in this context~\cite{sfermion2,p3b3,devereaux}. 
Thus, our focus and main question addressed are timely:  
{\it can we find an effective model for $d$-wave superconductivity using only the oxygens of a CuO$_2$
lattice, namely only the atoms placed at the bonds of the said square lattice?}  Moreover, in 
searching for the most compact in size form for this pairing operator we will address, as a bonus, 
the recent photoemission results in the cuprates that unveiled
very small Cooper pairs, at least in the antinodal directions~\cite{smallpairs}. Our overarching goal 
can be framed similarly as early studies within one-orbital models that attempted to construct
quasiparticle operators with a larger quasiparticle weight $Z$ that those of the usual bare operators (i.e. better
"antennas"),  and thus derive pairing operators that could produce stronger signals in computational 
studies~\cite{sch-dag}.

This paper is organized as follows: in Section~\ref{3band} the models traditionally used to study 
the cuprates, as well as the $d$-wave pairing operators previously investigated, are discussed. 
Our new minimal $d$-wave pairing operator in the CuO$_2$ planes is introduced in 
Section~\ref{minimal}, while in Section~\ref{time} additional $d$-wave pairing operators, 
including the more standard extended ones, 
are deduced by calculating the time evolution of the minimal operator. Both the minimal and some 
extended pairing operators are studied at the mean-field level in Section~\ref{MF} and a simple 
effective model is introduced in Section~\ref{effective}. Section~\ref{conclu} is devoted to our conclusions.

\section{Models and Previously Used $d$-wave Pairing Operators for Cuprates }\label{3band}

\begin{figure}[thbp]
\begin{center}
\includegraphics[width=7cm,clip,angle=0]{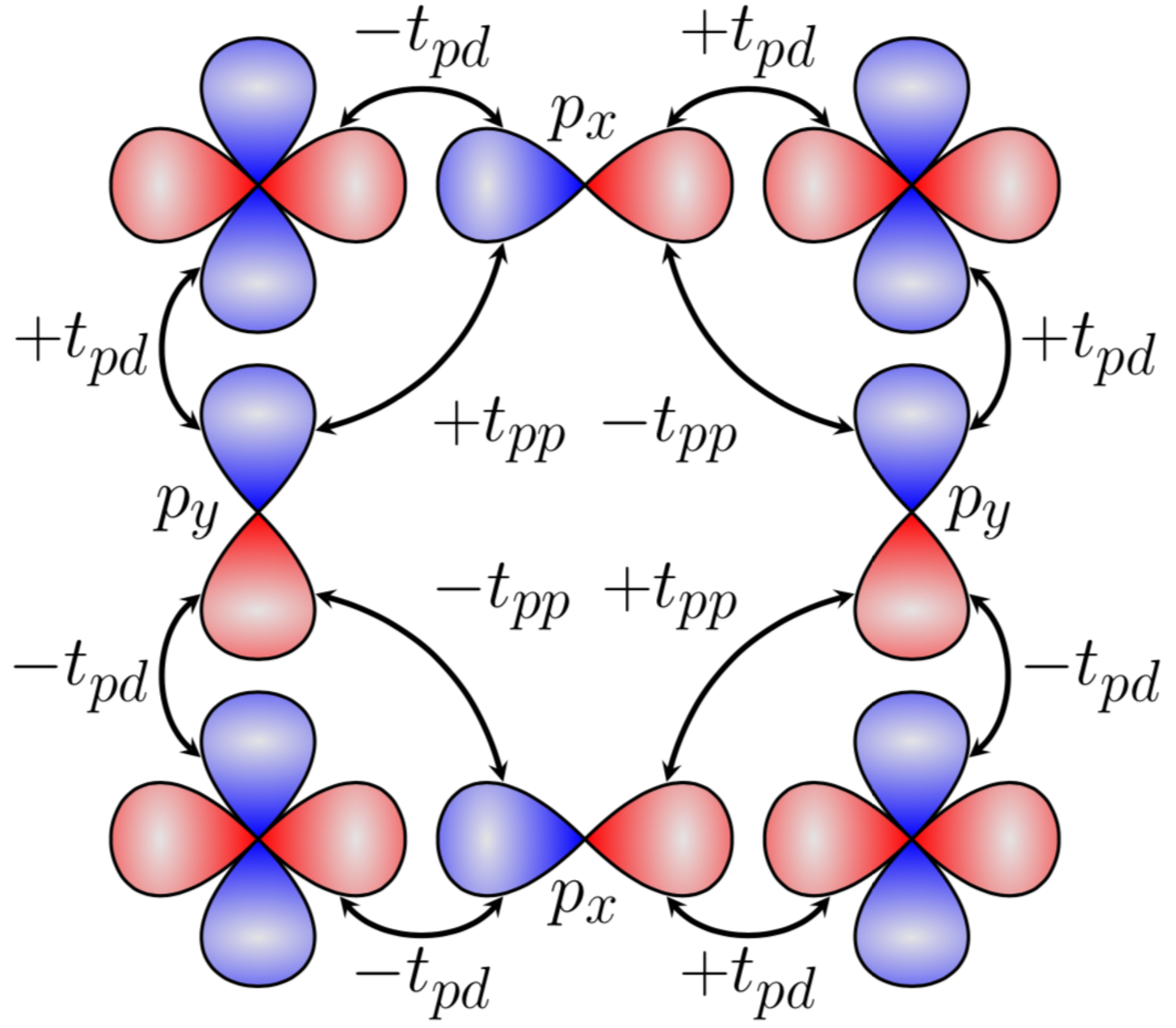}
\vskip -0.3cm
\caption{(color online) Schematic drawing of the Cu $d_{x^2-y^2}$ orbitals at the copper sites of the square lattice,
with the sign convention indicated by the colors (red for + and blue for -). The oxygen $p_{\sigma}$ orbitals 
with their corresponding sign convention are also shown, located at the Cu-O-Cu bonds.
The sign convention for the $t_{pd}$ and $t_{pp}$ hoppings is also presented.} 
\vskip -0.4cm
\label{cuo2fig}
\end{center}
\end{figure}

It is widely accepted that a realistic model to describe CuO$_2$ planes 
is a three-orbital Hubbard model that includes the $d_{x^2-y^2}$ orbitals
at the coppers and the $p_{\sigma}$ orbitals at the oxygens at a distance $\hat\mu/2$ from the coppers 
(lattice constant units), with $\hat\mu=x$ or $y$~\cite{emery} i.e. along the two directions. The 
Hamiltonian is
\begin{equation}
H_{\rm 3BH}= H_{\rm TB}+ H_{\rm int},
\label{3bh}
\end{equation}
\noindent where 
\begin{equation}\begin{split}
H_{\rm TB} = -t_{pd}\sum_{{\bf i},\mu,\sigma}\alpha_{{\bf i},\mu}(p^{\dagger}_{{\bf i}+{\hat\mu\over{2}},\mu,\sigma}d_{{\bf i},\sigma}+ h.c.)
-\\
t_{pp}\sum_{{\bf i},\langle\mu,\nu\rangle,\sigma}\alpha'_{{\bf i},\mu,\nu}[p^{\dagger}_{{\bf i}+{\hat\mu\over{2}},\mu,\sigma}(p_{{\bf i}+{\hat\nu\over{2}},\nu,\sigma}+p_{{\bf i}-{\hat\nu\over{2}},\nu,\sigma})+ h.c.]\\
+\epsilon_d\sum_{{\bf i}}n^d_{{\bf i}}+\epsilon_p\sum_{{\bf i},\mu}n^p_{{\bf i}+{\hat\mu\over{2}}}+\mu_e\sum_{{\bf i},\mu}(n^p_{{\bf i}+{\hat\mu\over{2}}}+n^d_{\bf i}),
\label{htb}
\end{split}\end{equation}
\noindent and
\begin{equation}
\begin{split}
H_{\rm int} = U_{d}\sum_{{\bf i}}n^d_{\bf i,\uparrow}n^d_{\bf i,\downarrow}
+U_{p}\sum_{{\bf i},\mu,\sigma}n^p_{{\bf i}+{\hat\mu\over{2}},\uparrow}n^p_{{\bf i}+{\hat\mu\over{2}},\downarrow}.
\label{int}
\end{split}
\end{equation}
\noindent The operator $d^{\dagger}_{{\bf i},\sigma}$ creates an electron with spin $\sigma$ 
at site ${\bf i}$ of the copper square lattice, while
$p^{\dagger}_{{\bf i}+{\hat\mu\over{2}},\mu,\sigma}$ creates an electron with spin $\sigma$ 
at orbital $p_{\mu}$, where $\mu=x$ or $y$, for the oxygen located at
${\bf i}+{\hat\mu\over{2}}$. The hopping amplitudes $t_{pd}$ and $t_{pp}$ 
correspond to the hybridizations between nearest-neighbors 
Cu-O and O-O, respectively, and $\langle\mu,\nu\rangle$ indicate O-O pairs connected by $t_{pp}$ as shown in
Fig.~\ref{cuo2fig}. $n^p_{{\bf i}+{\hat\mu\over{2}},\sigma}$ ($n^d_{{\bf i},\sigma}$) 
is the number operator for $p$ ($d$) electrons with spin $\sigma$, and
$\epsilon_d$ and $\epsilon_p$ are the on-site energies at the Cu and O sites, respectively. 
The Coulomb repulsion between two electrons at the same site and 
orbital is $U_d$ ($U_p$) for $d$ ($p$) orbitals. The signs
of the Cu-O and O-O hoppings due to the symmetries of the orbitals is included in 
the parameters $\alpha_{{\bf i},\mu}$ and $\alpha'_{{\bf i},\mu,\nu}$ and follow 
the convention shown in Fig.~\ref{cuo2fig}. Finally, $\mu_e$ is the electron chemical potential.
The hopping parameters are those much used for the cuprates i.e. $t_{pd}=1.3$~eV and $t_{pp}=0.65$~eV, 
on-site energy $\epsilon_p=-3.6$~eV~\cite{hybertsen}, and 
$\Delta_{CT}=\epsilon_d-\epsilon_p$ which is positive ($\epsilon_d=0$)~\cite{sfermion1} is the charge-transfer gap.

\subsection{Single-orbital $d$-wave operators}

Because experiments 
indicate that the Fermi surface of the cuprates is determined by a single band~\cite{sb1,sb2,sb3,sb4}, 
and theoretically a mapping of the three-band Hubbard model to the 
$t-J$ Hamiltonian can be obtained via Zhang-Rice singlets~\cite{ZR}, using only one band is appealing. 
In fact, due to their relative 
simplicity, the study of single-orbital models has prevailed in the cuprates. 
As a result, the simplest pairing operators with $d$-wave symmetry are extended in the sense that they 
involve nearest-neighbor Cu sites~\cite{elbioreview,Doug}, without the oxygens in between. 
Zhang and Rice studied
the addition of one hole in an undoped three-orbital Hubbard model using a CuO$_4$ cluster but 
neglecting the O-O hopping. They found that the hole occupies a symmetric linear 
combination involving the four
O's around a Cu, and forms a spin singlet together with the hole in the central Cu~\cite{ZR}. 
They also showed that 
the energy of the small cluster with two extra holes in the O orbitals was higher than 
the energy of two separated O holes. The next step was to construct Wannier functions combining 
the single-cluster symmetric single-hole plaquette states and obtain the effective
single-orbital low-energy model, leading to  the $t-J$ model. As discussed earlier, 
the simplest $d$-wave pairing operator in the $t-J$ (and one-orbital Hubbard) modes involves nearest-neighbor 
sites and has the well-known form:     
\begin{equation}
\Delta^{\dagger}_D({\bf j})=\sum_{\mu,\sigma}f(\sigma)\gamma_{\mu}c^{\dagger}_{{\bf j}+\hat\mu,\sigma}c^{\dagger}_{{\bf j},-\sigma},
\label{DtJ}
\end{equation}
\noindent where $c^{\dagger}_{{\bf j},\sigma}$ creates an electron with spin $\sigma$ at site ${\bf j}$ 
of the Cu square lattice [see panel (a) of Fig.~\ref{pairing1}], $\gamma_{\mu}$=1 (-1) for 
$\mu=\pm x$ ($\pm y$) and $f(\sigma)=1 (-1)$ if $\sigma=\uparrow (\downarrow)$. 

\subsection{Three-orbital extended $d$-wave operators}

\begin{figure}[thbp]
\begin{center}
\includegraphics[width=7cm,clip,angle=0]{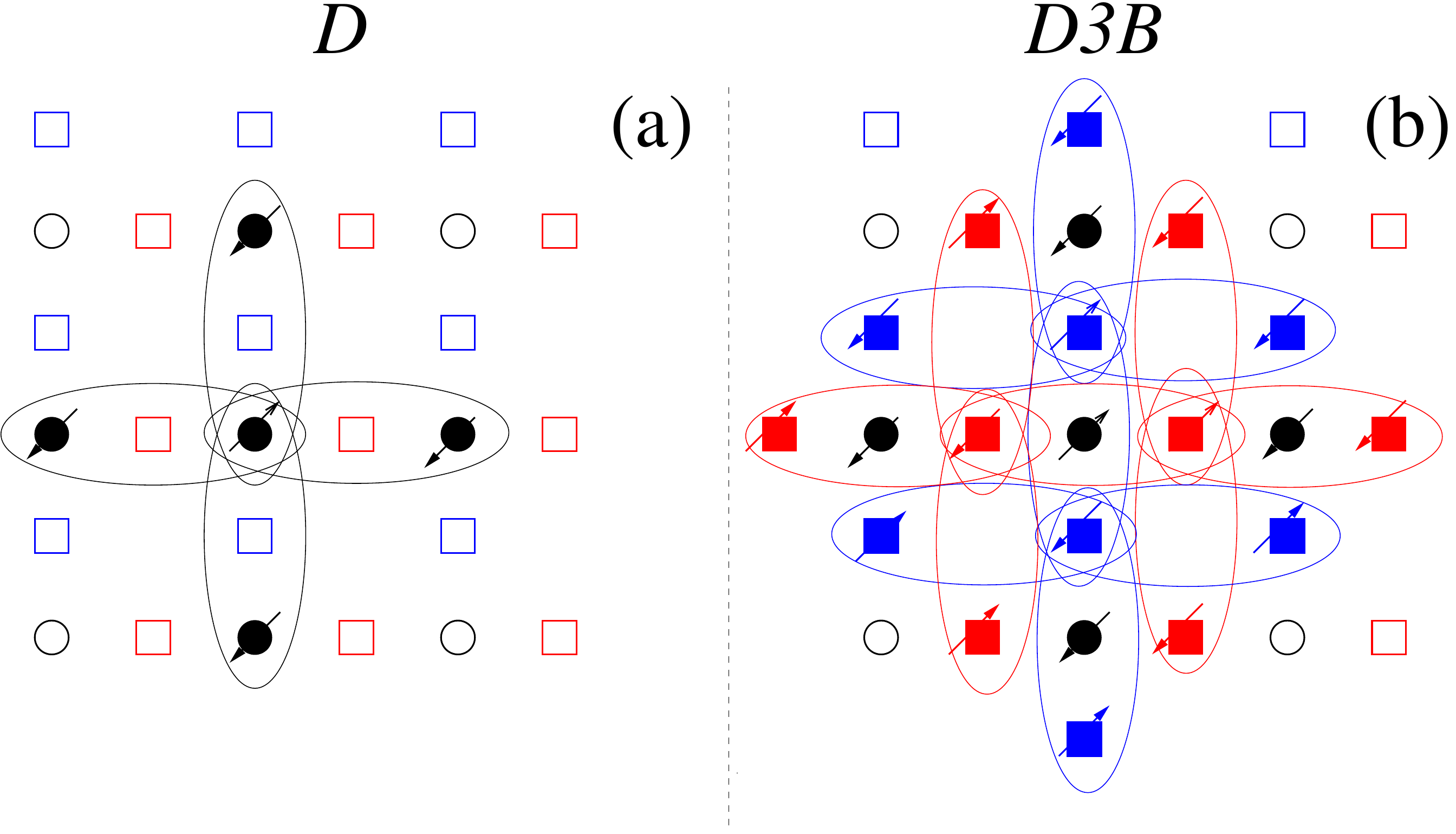}
\vskip -0.3cm
\caption{(color online) Schematic drawing of previously used $d$-wave pairing operators in the CuO$_2$ planes. 
Circles indicate the Cu $d$-orbitals sites while red (blue) squares indicate the O $p_{\rm x}$ 
($p_{\rm y}$) orbitals sites. Filled
symbols indicate atoms where the particles forming the Cooper pairs are located and 
arrows indicate the spin of the electrons/holes in the pair. (a) In single-orbital approximations 
to the CuO$_2$ planes the Cooper pairs are assumed to be primarily located in nearest-neighbor Cu 
sites via the operator $\Delta^{\dagger}_D$. (b) In the three-orbital Hubbard model the $d$-wave pairing 
operator $\Delta^{\dagger}_{D3B}$ adds Cooper-pairs involving $p_{\rm x}$ and 
$p_{\rm y}$ orbitals, in addition to the Cu 
orbitals as in (a) not shown in this panel for clarity. The individual Cooper pairs are encircled with ellipses.
The relative phases are positive along $x$ and negative along $y$.} 
\vskip -0.4cm
\label{pairing1}
\end{center}
\end{figure}

Note that the empty sites (holes) in the effective 
$t-J$ model contain Zhang-Rice singlets (ZRS) which means that the components of the 
Cooper pair in Eq.~(\ref{DtJ}) are created on top of the ZRS. The first numerical calculations studying 
pairing were performed in single-orbital models~\cite{elbioreview} and when, later on, 
pairing was numerically evaluated in three-orbital Hubbard models, the pairing operators 
used~\cite{p3b1,p3b2,p3b3} were straightforward generalizations of Eq.~(\ref{DtJ}) 
[see panel (b) of Fig.~\ref{pairing1}] involving several sites, such as 
\begin{equation}
\begin{split}
\Delta^{\dagger}_{D3B}({\bf j})=\sum_{\mu,\sigma}f(\sigma)\gamma_{\mu}[d^{\dagger}_{{\bf j}+\hat\mu,\sigma}d^{\dagger}_{{\bf j},-\sigma}+\\
p^{\dagger}_{{\bf j}+\hat\mu+x/2,x,\sigma}p^{\dagger}_{{\bf j}+x/2,x,-\sigma}+\\
p^{\dagger}_{{\bf j}+\hat\mu+y/2,y,\sigma}p^{\dagger}_{{\bf j}+y/2,y,-\sigma}].
\label{D3B}
\end{split}
\end{equation}
This operator creates electrons that form intra-orbital pairs whose $d$-wave symmetry 
is determined by $\gamma_{\mu}$. It considers that Cooper pairs are formed 
by one electron (or hole) in a Cu and another in its
neighboring Cu atoms, and similarly for electrons (or holes) in the $p$-orbitals. It is in
this sense that this operator is intra-orbital: the pair terms involve either Cu or O.
In real space the minimum pair created by the 
pairing operators in Eqs.~\ref{DtJ} and \ref{D3B} 
involves five lattice sites, in a single-orbital model context, 
or several unit cells (21 Cu and O sites) for the three-orbital case.
Such extended pairing operators [see panels (a) and (b) of Fig.~\ref{pairing1}] 
appear at odds with the recent experimental 
results of Ref.~\cite{smallpairs} where the observed pairs have a minimum real-space extension 
of the order of the lattice constant along the antinodal direction. 
This photoemission experiment offers motivation to investigate if in 
the CuO$_2$ planes it is possible to construct a more local $d$-wave pairing 
operator involving far less sites and ideally just one unit cell. 

\section{Minimal $d$-wave pairing operator}\label{minimal} 

As explained, in undoped systems and in the Zhang-Rice approximation 
a doped hole is placed at an oxygen and forms a ZRS 
with the hole at a copper. A second doped hole is expected to form another ZRS
with a different Cu. The one-orbital pairing operator in Eq.~(\ref{DtJ}) can 
only involve electrons at two neighboring Cu sites, each with its own ZRS. 
However, in the three-orbital Hubbard model formulation 
there is no clear relation between the pairing operator  and the two neighboring ZRS. 
Equation (\ref{D3B}) just considers that the minimal
Cooper pair can be formed by fermions at a $d$ ($p_{\sigma}$) orbital and at the four nearest-neighbor Cu (O) atoms, 
thus involving five unit cells, and many sites.

As discussed above, 
Zhang and Rice found out that it would be unlikely that two holes would share the O orbitals of 
one single plaquette. However, calculations including $t_{pp}$ hopping and the $p-d$ Coulomb 
repulsion, both 
neglected in the ZRS derivation, indicated that an effective attraction between holes in the oxygens of a single plaquette may develop~\cite{varma,cristian}. Thus, the possibility 
that two holes could form a pair in 
the O orbitals in a single plaquette deserves to be explored.
 
\begin{figure}[thbp]
\begin{center}
\subfigure{\includegraphics[trim = 8mm 5mm 4mm 5mm,width=0.45\textwidth,angle=0]{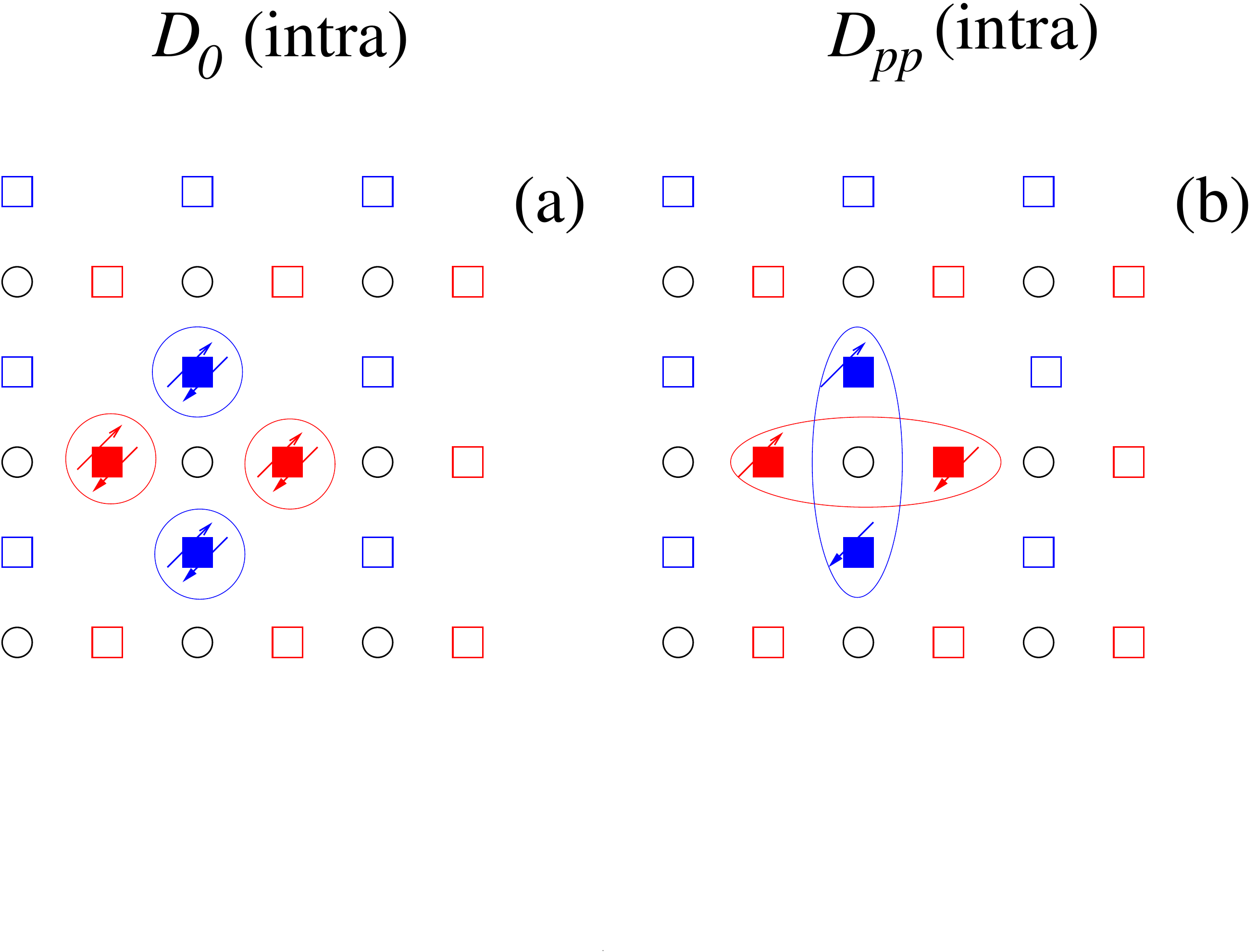}}
\subfigure{\includegraphics[trim = 8mm 5mm 4mm 5mm,width=0.45\textwidth,angle=0]{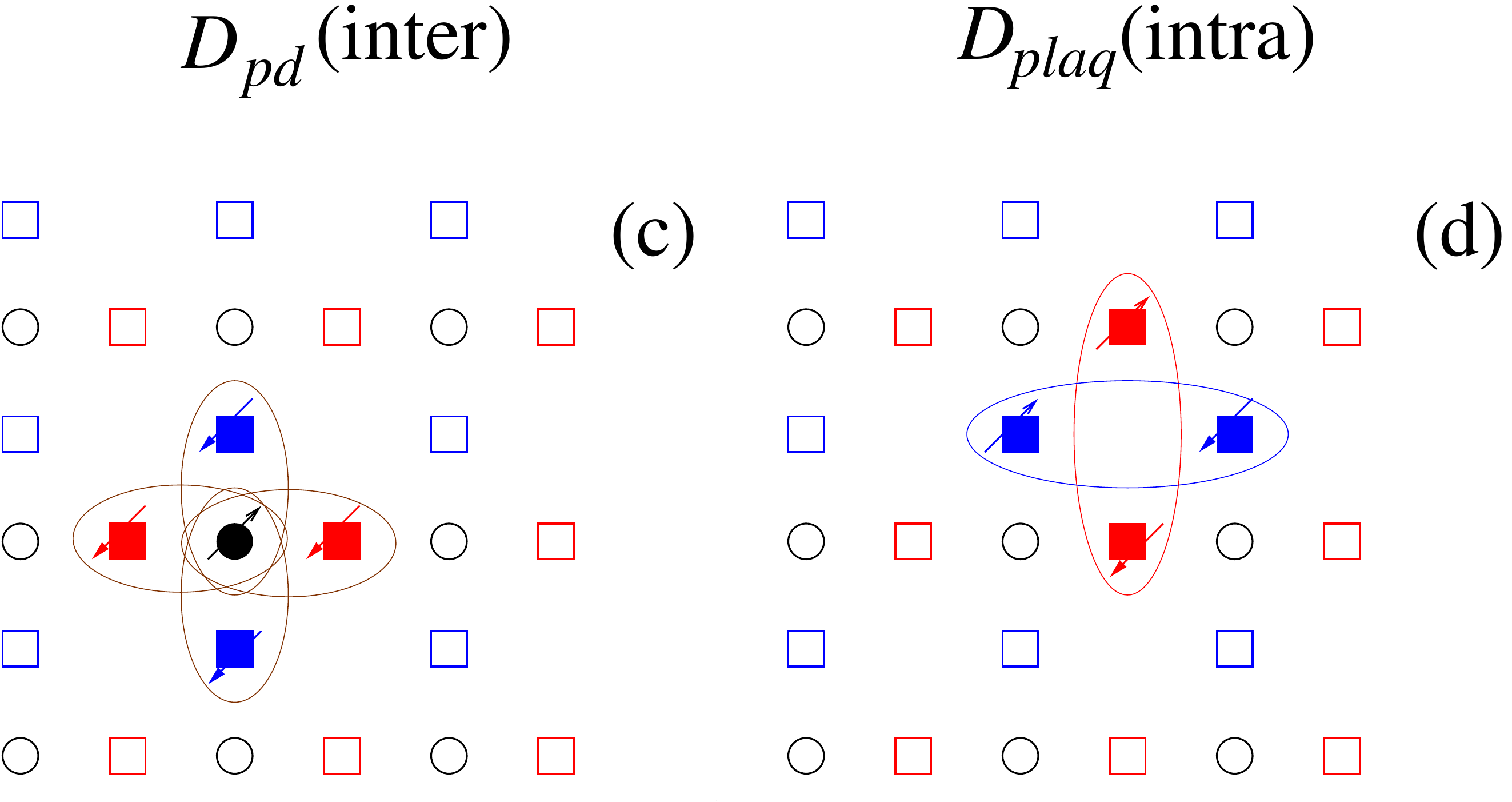}}
\vskip -0.2cm
\caption{(color online) Schematic drawing of the minimal intra-unit cell $d$-wave 
pairing operators introduced here
for the CuO$_2$ planes, that were not explored before in three-orbital Hubbard models to our knowledge. 
Circles indicate the Cu $d$-orbitals while the red (blue) squares 
indicate the $p_{\rm x}$ ($p_{\rm y}$) 
orbitals at the O atoms. Filled
symbols indicate sites where the particles forming the Cooper pairs are located 
and arrows indicate the spin of the electrons/holes in the pair. 
(a) On-site intra-orbital (i.e. same oxygen) $d$-wave operator 
$\Delta^{\dagger}_{D0}$, defined in Eq.~(6). Here the two members of the Cooper 
pair are at the same oxygen, linearly combined involving the four possible oxygens. 
(b) More extended nearest-neighbor intra-orbital  $d$-wave operator $\Delta^{\dagger}_{Dpp}$ where the Cooper
pair is formed by two electrons in the same $p_{\sigma}$ orbital, either $x$ or $y$, at a distance
of one lattice spacing 
forming a spin singlet, and linearly
combining the vertical and horizontal directions to form a $d$-wave operator. 
(c) Inter-orbital ($dp$) $d$-wave operator $\Delta^{\dagger}_{Dpd}$ with 
pairs involving a particle at the central Cu
and the other at a neighboring O, linearly combined to form a $d$-wave. 
(d) Plaquette $d$-wave intraorbital pairing operator $\Delta^{\dagger}_{Dplaq}$ in the CuO$_2$ plane. In (a-d), 
the Cooper pairs are encircled with ellipses. 
Relative phases are positive along $x$ and negative along $y$, leading to a $d$-wave.} 
\vskip -0.4cm
\label{pairing2}
\end{center}
\end{figure}

First, we will construct an on-site $d$-wave pairing operator which considers only {\it doubly occupied} 
O sites [see panel (a) in Fig.~\ref{pairing2}] in analogy with the on-site attractive $s$-wave pairing operator.
It has the form
\begin{equation}
\begin{split}
\Delta^{\dagger}_{D0}({\bf j})=
 {1\over{2}}\sum_{\mu,\sigma}f(\sigma)\gamma_{\mu}p^{\dagger}_{{\bf j}+\hat\mu/2,\mu,\sigma}p^{\dagger}_{{\bf j}+\hat\mu/2,\mu,-\sigma}=\\
\sum_{\mu}\gamma_{\mu}p^{\dagger}_{{\bf j}+\hat\mu/2,\mu,\uparrow}p^{\dagger}_{{\bf j}+\hat\mu/2,\mu,\downarrow}.
\label{dw3bs}
\end{split}
\end{equation}
\noindent Although the operator involves doubly-occupied sites, each one apparently $s$-wave, 
since the operator involves four oxygens around the same copper, 
a linear combination can be made that renders the full operator $d$-wave. 

\section{Time-Evolution of the pairing operator}\label{time} 

In previous literature~\cite{zhang,you} a relationship between the on-site 
and the extended $s$-wave pairing operators in the single-orbital Hubbard model was 
obtained by calculating the time evolution of the on-site pairing operator. Following similar 
steps we can now calculate the time evolution of the on-site minimal $d$-wave pairing operator
$\Delta^{\dagger}_{D0}$ proposed in Eq.~(\ref{dw3bs}) for the three-orbital Hubbard model Eq.~(\ref{3bh}). 
We found that 
\begin{equation}
\begin{split}
-i{d\Delta^{\dagger}_{D0}\over{dt}}=[H_{3BH},\Delta^{\dagger}_{D0}]=\\
2(\epsilon_p-\mu_e)\Delta^{\dagger}_{D0}-U_p\Delta^{\dagger}_{D0}-t_{pd}\Delta^{\dagger}_{Dpd},
\label{com}
\end{split}
\end{equation}
\noindent where $\Delta^{\dagger}_{Dpd}$ is another $d$-wave pairing 
operator defined in one unit-cell CuO$_2$. $\Delta^{\dagger}_{Dpd}$ forms Cooper pairs 
with one fermion at a Cu and the other in an antisymmetric linear combination of the p$_{\sigma}$ 
orbitals in its four nearest-neighboring O's [panel (c) of Fig.~\ref{pairing2}] and it is given by
\begin{equation}
\Delta^{\dagger}_{Dpd}({\bf j})=
 \sum_{\mu,\sigma}f_{\sigma}\gamma_{\mu}\alpha_{{\bf j},\mu}d^{\dagger}_{{\bf j},\sigma}p^{\dagger}_{{\bf j}+\hat\mu/2,\mu,-\sigma}.
\label{dwpd}
\end{equation}

Since in the ground state the average value of the pairing operators is time-independent, from Eq.~(\ref{com}) we see that the average values of the two pairing 
operators must be related. In addition, by evaluating the time evolution of the new
inter-orbital minimal pairing operator, $\Delta^{\dagger}_{Dpd}$, more extended intra- and inter-orbital pairing operators with $d$-wave symmetry are obtained. 
For example, from the commutator between $\Delta^{\dagger}_{Dpd}$ and the $t_{pd}$ 
hopping term in $H_{\rm 3BH}$ we obtain the nearest-neighbor $d$-orbital pairing operator 
in Eq.~(\ref{DtJ}) depicted in panel (a) of Fig.~\ref{pairing1} and an additional 
intra-orbital pairing operator given by
\begin{equation}
\begin{split}
\Delta^{\dagger}_{Dpp}({\bf j})=
 (p^{\dagger}_{{\bf j}+x/2,x,\uparrow}p^{\dagger}_{{\bf j}-x/2,x,\downarrow}-p^{\dagger}_{{\bf j}+x/2,x,\downarrow}p^{\dagger}_{{\bf j}-x/2,x,\uparrow})-\\
(p^{\dagger}_{{\bf j}+y/2,y,\uparrow}p^{\dagger}_{{\bf j}-y/2,y,\downarrow}-p^{\dagger}_{{\bf j}+y/2,y,\downarrow}p^{\dagger}_{{\bf j}-y/2,y,\uparrow})=\\
 \sum_{\mu,\sigma}f_{\sigma}\gamma_{\mu}p^{\dagger}_{{\bf j}+\hat\mu/2,\mu,\sigma}p^{\dagger}_{{\bf j}-\hat\mu/2,\mu,-\sigma},
\label{dw3bl}
\end{split}
\end{equation}
\noindent 
which is another $B_{1g}$ intra-orbital 
pairing operator with the two particles located in the same orbital but 
at different oxygens [see panel (b) in Fig.~\ref{pairing2}]
and it is analogous to the extended, nearest-neighbor, $s$-wave operator defined in the context of the cuprates~\cite{elbioreview,zhang}. 
In addition, the commutator between $\Delta^{\dagger}_{Dpd}$ 
and the $t_{pp}$ hopping term in $H_{\rm 3BH}$ leads to an extended version of $\Delta^{\dagger}_{Dpd}$ that forms pairs with one fermion on a 
$d$ orbital at site ${\bf r}$ and the other at orbital $p_x$ ($p_y$) at distance ${\bf r}+y+x/2$ (${\bf r}+x+y/2$) and symmetrical points.
The commutator of this extended operator with the $t_{pd}$ hopping term in $H_{\rm 3BH}$ finally leads to  
$p$-orbital pairing operators that combine fermions in $p_{\rm x}$ ($p_{\rm y}$) orbitals 
along the $y$ ($x$)
direction which are the plaquette pairing operators mentioned in Ref.~\cite{smallpairs} 
and shown in panel (d) of Fig.~\ref{pairing2}. They are given by

\begin{equation}
\begin{split}
\Delta^{\dagger}_{Dplaq}({\bf j})=
 (p^{\dagger}_{{\bf j}+x/2,x,\uparrow}p^{\dagger}_{{\bf j}+y+x/2,x,\downarrow}-\\
p^{\dagger}_{{\bf j}+x/2,x,\downarrow}p^{\dagger}_{{\bf j}+y+x/2,x,\uparrow})-\\
(p^{\dagger}_{{\bf j}+y/2,y,\uparrow}p^{\dagger}_{{\bf j}+x+y/2,y,\downarrow}-\\
p^{\dagger}_{{\bf j}+y/2,y,\downarrow}p^{\dagger}_{{\bf j}+x+y/2,y,\uparrow})=\\
 \sum_{\mu,\sigma}f_{\sigma}\gamma_{\mu}p^{\dagger}_{{\bf j}+\hat\mu/2,\mu,\sigma}p^{\dagger}_{{\bf j}+\bar\mu+\hat\mu/2,\mu,-\sigma},
\label{dwplaq}
\end{split}
\end{equation}
\noindent where $\bar\mu=x$ ($y$) if $\mu=y$ ($x$).
We also notice that the $p$ contribution 
in the standard $d$-wave pairing operator in Eq.~(\ref{D3B}) results from a combination of the 
intra-orbital $p$ pairing operators operators $\Delta^{\dagger}_{Dplaq}$ and $\Delta^{\dagger}_{Dpp}$. 

The relationships between the minimal and the compact, but more extended, $d$-wave pairing 
operators in Fig.~\ref{pairing2} deduced from the time-evolution calculations suggest that
if the Hamiltonian indeed has a superconducting ground state with $d$-wave symmetry we would expect 
that $all$ the pairing operators with that symmetry will develop long-range order simultaneously. In
practice we expect the long-range  behavior of local operators 
to be easier to study in the finite, often small, clusters accessible to numerical studies.
For this reason, we will focus on the newly introduced $d$-wave pairing 
operators shown in Fig.~\ref{pairing2} and we will compare them with the traditional ones presented in Fig.~\ref{pairing1}.

\section{Effective model for $d$-wave pairing}\label{MF}

To show explicitly that the pairing operators in Eq.~(\ref{dw3bs}), Eq.~(\ref{dwpd}), Eq.~(\ref{dw3bl}), and Eq.~(\ref{dwplaq}) 
indeed lead to $d$-wave superconductors we will study the phenomenological Hamiltonian given by
\begin{equation}
H_{\rm 3BDW}= H_{\rm TB}+ H_{\rm int},
\label{DW}
\end{equation}
\noindent where the tight-binding term is the canonical 
of the three-orbital Hubbard model for cuprates [Eq.~(\ref{htb})] and
the interacting portion of the Hamiltonian for the on-site same-oxygen pairing, as in $D_0$, is given by
\begin{equation}
\begin{split}
H^{(0)}_{\rm int}=
-\sum_{{\bf j},\mu,\sigma}\gamma_{\mu}f(\sigma)[p^{\dagger}_{{\bf j}+\hat\mu/2,\mu,\sigma}p^{\dagger}_{{\bf j}+\hat\mu/2,\mu,-\sigma}\Delta+\\
\Delta^*p_{{\bf j}-\hat\mu/2,\mu,-\sigma}p_{{\bf j}-\hat\mu/2,\mu,\sigma}].
\label{hintreals}
\end{split}
\end{equation}
$\Delta$ and $\Delta^*$ are parameters that determine the strength of the superconducting condensate and they contain also the
attractive coupling $V$ usually employed in these phenomenological models. 
For the case of the inter-orbital 
extended pairing $\Delta^{\dagger}_{Dpd}$
the interaction term is given by
\begin{equation}
\begin{split}
H^{(pd)}_{\rm int}=
-\sum_{{\bf j},\mu,\sigma}\gamma_{\mu}f(\sigma)\alpha_{{\bf j},\mu}[d^{\dagger}_{{\bf j},\sigma}p^{\dagger}_{{\bf j}-\mu/2,\mu,-\sigma}\Delta+\\
\Delta^*p_{{\bf j}-\mu/2,\mu,-\sigma}d_{{\bf j},\sigma}].
\label{hintrealpd}
\end{split}
\end{equation}
For the intra-orbital 
extended pairing $\Delta^{\dagger}_{Dpp}$
the interaction term is given by

\begin{equation}
\begin{split}
H^{(pp)}_{\rm int}=
-\sum_{{\bf j},\mu,\sigma}\gamma_{\mu}f(\sigma)[p^{\dagger}_{{\bf j}+\hat\mu/2,\mu,\sigma}p^{\dagger}_{{\bf j}-\hat\mu/2,\mu,-\sigma}\Delta+\\
\Delta^*p_{{\bf j}-\hat\mu/2,\mu,-\sigma}p_{{\bf j}+\hat\mu/2,\mu,\sigma}],
\label{hintreale}
\end{split}
\end{equation}
\noindent while for the plaquette operator $\Delta^{\dagger}_{Dplaq}$ the interaction is given by
\begin{equation}
\begin{split}
H^{(plaq)}_{\rm int}=
-\sum_{{\bf j},\mu,\sigma}\gamma_{\mu}f(\sigma)[p^{\dagger}_{{\bf j}+\hat\mu/2,\bar\mu,\sigma}p^{\dagger}_{{\bf j}-\hat\mu/2,\bar\mu,-\sigma}\Delta+\\
\Delta^*p_{{\bf j}-\hat\mu/2,\bar\mu,-\sigma}p_{{\bf j}+\hat\mu/2,\bar\mu,\sigma}].
\label{hintrealp}
\end{split}
\end{equation}

\subsection{Mean-Field analysis}\label{MF1} 

In this section we perfom a canonical mean-field analysis of the effective pairing models now
using the more compact $d$-wave operators introduced here. As usual, via a Fourier transform 
we can work in momentum space which is more convenient. Thus, $H_{\rm TB}$ can be written as 
\begin{equation}
H_{\rm TB}( {\bf k})=\sum_{ {\bf k},\sigma}\Phi^{\dagger}_{ {\bf k},\sigma}\xi_{ {\bf k}}
\Phi_{ {\bf k},\sigma},
\label{8}
\end{equation}
where  
$\Phi^{\dagger}_{ {\bf k},\sigma}=(p^{\dagger}_{x}({\bf k}),p^{\dagger}_{y}({\bf k}),d^{\dagger}({\bf k}))_{\sigma}$ and
\begin{equation}
\xi_{\bf k}=
 \left(\begin{array}{ccc}
\epsilon_p    & -4t_{pp}s_xs_y & -2it_{pd}s_x  \\
-4t_{pp}s_xs_y & \epsilon_p & -2it_{pd}s_y  \\
2it_{pd}s_x & 2it_{pd}s_y  & 0
\end{array} \right),
\label{15}
\end{equation}
\noindent where $s_i$ indicates $\sin (k_i/2)$ with $i=x$ or $y$.

\begin{figure}[thbp]
\begin{center}
\includegraphics[width=8.5cm,clip,angle=0]{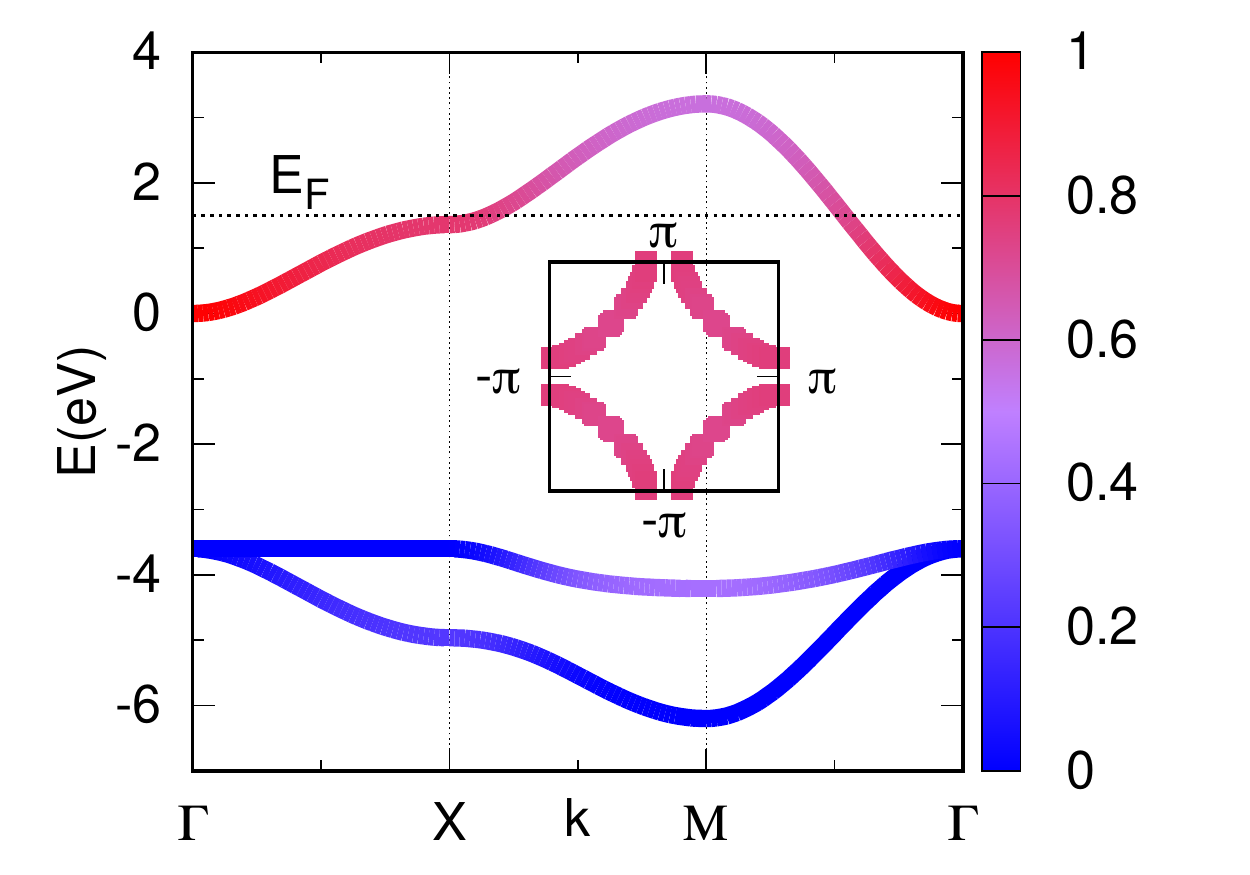}
\vskip -0.3cm
\caption{(color online) Band dispersion for the tight-binding term of the CuO$_2$ Hamiltonian. 
The orbital content is displayed with red (blue) indicating $d$ ($p$) character. The dashed
line indicates the position of the chemical potential (or Fermi level E$_{\rm F}$) 
at density $\langle n \rangle=5$ (undoped case). 
The Fermi surface at this density is in the inset. Colors indicate the 
orbital content of the bands, with
the palette on the right denoting the weight of the $d$ component (e.g. 1 means 
100\% copper $d$, and the oxygen weight is simply one minus the copper weight).} 
\vskip -0.4cm
\label{TBdisp}
\end{center}
\end{figure}

Note that in the electron representation the undoped case is characterized by 
one hole at the Cu and no holes at the O, which 
corresponds to a total of five electrons per CuO$_2$ unit-cell (the maximum possible electronic 
number in three orbitals is six). The orbital-resolved tight-binding bands along the 
$\Gamma-X-M-\Gamma$ path in the Brillouin zone calculated using a $100\times 100$ square lattice 
(with Cu's at the sites of the lattice) is in Fig.~\ref{TBdisp}. The dashed black line is 
the chemical potential $\mu_e$ for the important electronic density $\langle n\rangle=5$ and the 
corresponding Fermi surface is in the inset. An analysis of the orbital composition 
of each of the three bands, shown by the color palette in the figure, indicates that the 
top band is purely $d$ at the $\Gamma$ point and moving away from $\Gamma$ becomes hybridized 
with the $p$ orbitals such that its $d$ content 
becomes 78\% at X and 56\% at M. The two bottom bands have pure $p$ 
character at the Brillouin zone center. The middle band achieves 43\% $d$ character at M, while 
the lower band has 21\% $d$ character at X. Note that the
tight-binding Fermi surface, shown in the inset, has the qualitative form expected 
in the cuprates, both from the theory and experimental perspectives. However, its orbital content 
is only about 75\% $d$ in average, showing that the oxygen component
is not negligible even if only one band crosses the Fermi level.
  
Note that $\xi_{\bf k}$ can be written in terms of the $3 \times 3$ Gell'mann matrices~\cite{schiff} $\lambda_i$ for the cases 
$i=1$ to 8, while $\lambda_0$ is the $3\times 3$ identity (see Appendix for an explicit form of these matrices). This is useful in order to
highlight the symmetry of its different terms:
\begin{equation}
\xi_{ {\bf k}}={2\over{3}}\epsilon_p\lambda_0+{\sqrt{3}\over{3}}\epsilon_p\lambda_8+2t_{dp}(s_x\lambda_5+s_y\lambda_7)-
4t_{pp}s_xs_y\lambda_1.
\label{9}
\end{equation}

Since the CuO$_2$ planes transform as $D_{\rm 4h}$ and the Hamiltonian has to be invariant under the group operations, i.e., it has to 
transform as the $A_{1g}$ representation of the group, we notice 
that in Eq.~(\ref{9}), $\lambda_0$ and $\lambda_8$ transform 
like $A_{1g}$ while $\lambda_1$ transforms like $B_{2g}$, and $(\lambda_5,\lambda_7)$ transform 
like the two-dimensional representation $E_{g}$ since they are combined 
with $(s_x,s_y)$, which transforms according to $E_{g}$.

The interacting term of the Hamiltonian can be written in terms of a pairing matrix $P^{(0)}$ for the on-site same-oxygen case (Eq.~(\ref{hintreals})):
\begin{equation}
P^{(0)}_{\bf k}=
 \left(\begin{array}{ccc}
2\Delta & 0 & 0  \\
0 & -2\Delta  & 0  \\
0 & 0 & 0    
\end{array} \right),
\label{155}
\end{equation}
which can be written in terms of the $\lambda_i$ matrices as
\begin{equation}
P^{(0)}_{ {\bf k}}=2\Delta \lambda_3.
\label{99}
\end{equation}

For the extended pairing (Eq.~(\ref{hintreale})) the corresponding matrix $P^{(pp)}$ is given by
\begin{equation}
P^{(pp)}_{\bf k}=
 \left(\begin{array}{ccc}
2\Delta\cos(k_x) & 0 & 0  \\
0 & -2\Delta\cos(k_y)  & 0  \\
0 & 0 & 0    
\end{array} \right),
\label{1555}
\end{equation}
which can be written in terms of the $\lambda_i$ matrices as
\begin{equation}
\begin{split}
P^{(pp)}_{ {\bf k}}=2\Delta[{(\cos(k_x)-\cos(k_y))\over{3}}\lambda_0+{(\cos(k_x)+\cos(k_y))\over{2}}\lambda_3+\\
{\sqrt{3}(\cos(k_x)-\cos(k_y))\over{6}}\lambda_8.
\label{999}
\end{split}
\end{equation}

For the plaquette pairing operator (Eq.~(\ref{hintrealp})) the corresponding matrix $P^{(plaq)}$ is given by
\begin{equation}
P^{(plaq)}_{\bf k}=
 \left(\begin{array}{ccc}
2\Delta\cos(k_y) & 0 & 0  \\
0 & -2\Delta\cos(k_x)  & 0  \\
0 & 0 & 0    
\end{array} \right),
\label{1558}
\end{equation}
which can be written in terms of the $\lambda_i$ matrices as
\begin{equation}
\begin{split}
P^{(plaq)}_{ {\bf k}}=2\Delta[{(\cos(k_y)-\cos(k_x))\over{3}}\lambda_0+{(\cos(k_y)+\cos(k_x))\over{2}}\lambda_3+\\
{\sqrt{3}(\cos(k_y)-\cos(k_x))\over{6}}\lambda_8.
\label{9998}
\end{split}
\end{equation}

For the inter-orbital pairing operator $\Delta^{\dagger}_{Dpd}$ the corresponding matrix $P^{(pd)}$ (Eq.~(\ref{hintrealpd})) is given by
\begin{equation}
P^{(pd)}_{\bf k}=
 \left(\begin{array}{ccc}
0 & 0& 2\Delta i s_x  \\
0 & 0& -2\Delta i s_y  \\
2\Delta i s_x & -2\Delta i s_y & 0    
\end{array} \right),
\label{1559}
\end{equation}
which can be written in terms of the $\lambda_i$ matrices as
\begin{equation}
\begin{split}
P^{(pd)}_{ {\bf k}}=2i\Delta(s_x\lambda_4-s_y\lambda_6).
\label{9999}
\end{split}
\end{equation}

Thus, the effective interaction term can be constructed in terms of a spin-singlet pair operator that transforms according to the irreducible 
representation $B_{1g}$ of $D_{4h}$~\cite{3orbs}. In the case of $P^{(0)}$ the symmetry of the pairing term is 
given by the matrix $\lambda_3$, which transforms according to $B_{1g}$. For
$P^{(pp)}$ in Eq.~(\ref{999}) note that the terms that contain $\lambda_0$ 
and $\lambda_8$, which transform like $A_{1g}$, 
are multiplied by $\cos(k_x)-\cos(k_y)$, which transforms like $B_{1g}$, while the
term that contains $\lambda_3$, which transforms like $B_{1g}$, is 
multiplied by $\cos(k_x)+\cos(k_y)$, which transforms like $A_{1g}$. A similar analysis for $P^{(plaq)}$ 
in Eq.~(\ref{9998}) shows that
it also transforms like $B_{1g}$. The inter-orbital pairing 
operator also transforms as $B_{1g}$ because it combines $(s_x,s_y)$ with $(\lambda_4,\lambda_6)$
each transforming like $E_g$. Finally, for completeness, we present the pairing matrices for the traditional 
$d$-wave operator of the single and three-orbital Hubbard models
presented in Eq.~(\ref{DtJ}) and Eq.~(\ref{D3B}). 
For $\Delta^{\dagger}_D$ the corresponding matrix $P^{(D)}$ is
\begin{equation}
P^{(D)}_{\bf k}=
 \left(\begin{array}{ccc}
0 & 0 & 0  \\
0 & 0  & 0  \\
0 & 0 & 2\Delta[\cos(k_x)-\cos(k_y)]    
\end{array} \right),
\label{1548}
\end{equation}
which can be written in terms of the $\lambda_i$ matrices as
\begin{equation}
P^{(D)}_{ {\bf k}}=2\Delta[\cos(k_x)-\cos(k_y)](\lambda_0-\sqrt{3}\lambda_8),
\label{9978}
\end{equation}
\noindent and for $\Delta^{\dagger}_{D3B}$ the corresponding matrix $P^{(D3B)}$ is
\begin{equation}
P^{(D3B)}_{\bf k}=2\Delta[\cos(k_x)-\cos(k_y)]\\
 \left(\begin{array}{ccc}
1 & 0 & 0  \\
0 & 1  & 0  \\
0 & 0 & 1    
\end{array} \right),
\end{equation}
which can be written in terms of the $\lambda_i$ matrices as
\begin{equation}
\begin{split}
P^{(D3B)}_{ {\bf k}}=2\Delta[\cos(k_x)-\cos(k_y)]\lambda_0.
\label{9978}
\end{split}
\end{equation} 
In summary, for the canonical widely used operators the $B_{1g}$ symmetry is just 
directly given by the factor $\cos(k_x)-\cos(k_y)$, while for the new operators deducing the
$d$-wave character requires a careful analysis. 

Another way of verifying the $d$-wave symmetry of the proposed 
pairing operators is the calculation of the band structure 
via the resulting $6 \times 6$ Bogoliubov-de Gennes Hamiltonian given by

\begin{equation}
H_{\rm BdG}=\sum_{{\bf k}}\Psi^{\dagger}_{\bf k}H^{\rm MF}_{\bf k}\Psi_{\bf k},
\label{13}
\end{equation}
\noindent with the definitions
\begin{equation}
\Psi^{\dagger}_{\bf k}=(p^{\dagger}_{{\bf k},x,\uparrow},p^{\dagger}_{{\bf k},y,\uparrow},d^{\dagger}_{{\bf k},\uparrow},
p_{-{\bf k},x,\downarrow},p_{-{\bf k},y,\downarrow},d_{-{\bf k},\downarrow}),
\label{14}
\end{equation}
\noindent and
\begin{equation}
H^{\rm MF}_{\bf k}=
 \left(\begin{array}{cc}
(H_{\rm TB}({\bf k})-\mu_e\lambda_0) & P^{(\alpha)}( {\bf k}) \\
(P^{(\alpha)})^{\dagger}({\bf k}) & -(H_{\rm TB}({\bf k})-\mu_e\lambda_0)
\end{array} \right),
\label{15}
\end{equation}

\noindent where the label $\alpha$ takes the values $0$, ${pp}$, ${pd}$, ${plaq}$, $D$, or ${D3B}$, and we have included the chemical potential $\mu_e$ into the 
tight-binding term to ensure that the gap opens at the Fermi surface.

\begin{figure}[thbp]
\begin{center}
\subfigure{\includegraphics[trim = 8mm 5mm 4mm 5mm,width=0.45\textwidth,angle=0]{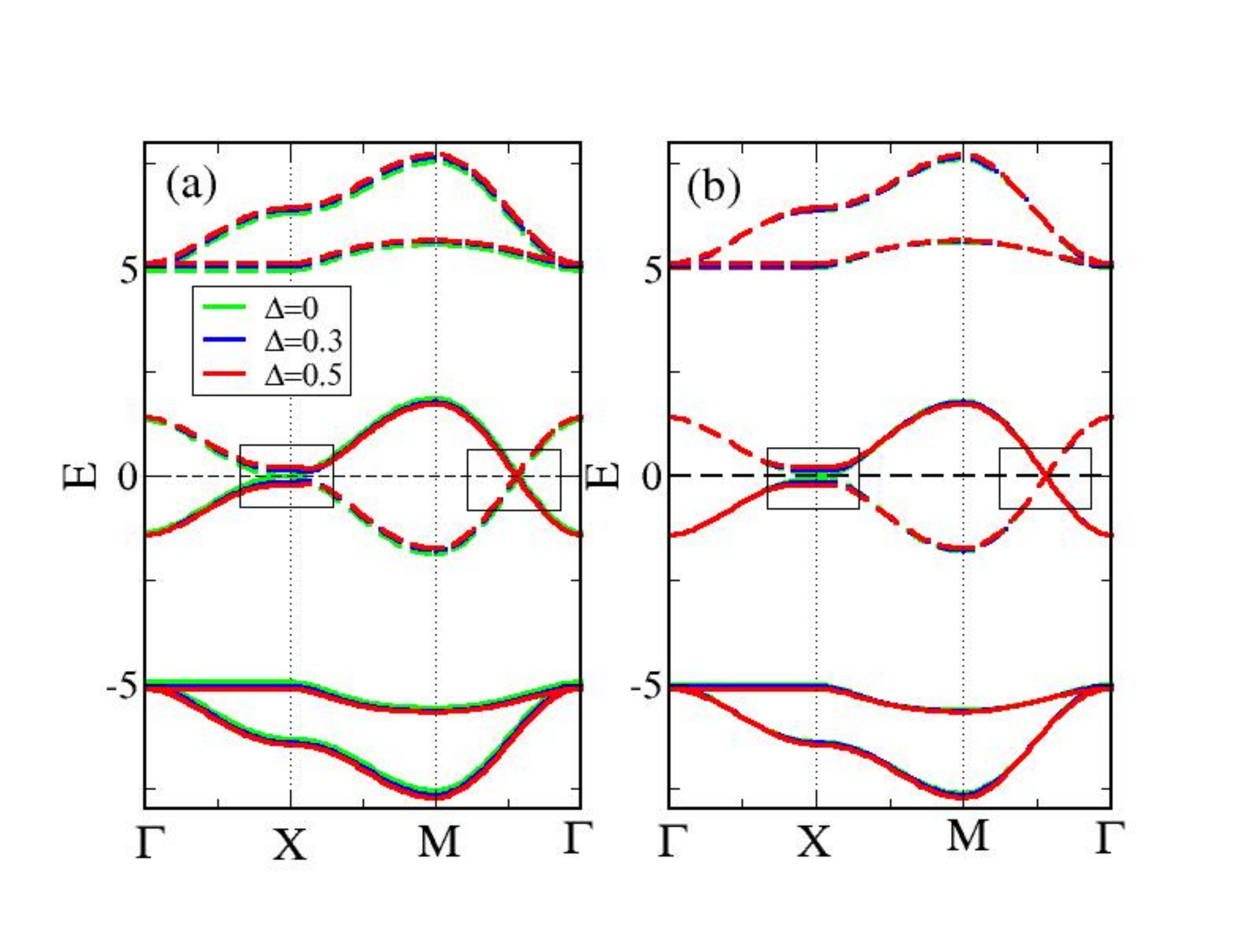}}
\subfigure{\includegraphics[trim = 8mm 5mm 4mm 5mm,width=0.45\textwidth,angle=0]{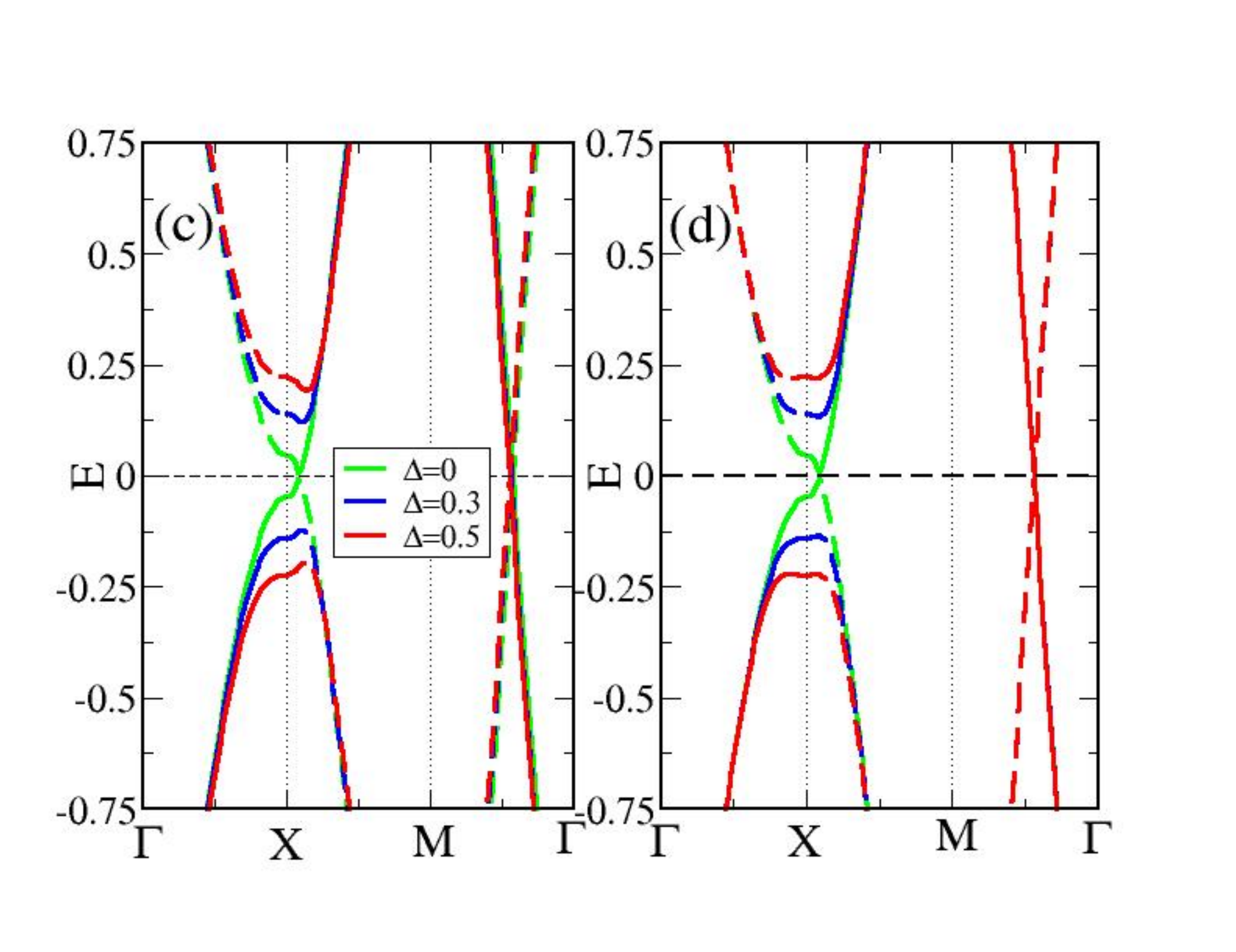}}
\vskip -0.3cm
\caption{(color online) Band dispersion for the mean-field Hamiltonians with $B_{1g}$ pairing. 
(a) corresponds to on-site $D_0$ and (b) to extended $D_{pp}$, for the indicated values 
of the pairing order parameter $\Delta$ at a density of 4.9 electrons per unit cell; (c) detail of the areas inside rectangular boxes in (a);
(d) detail of the areas inside rectangular boxes in (b). The dashed lines indicate ``shadow'' bands.} 
\vskip -0.4cm
\label{gap}
\end{center}
\end{figure}

Diagonalizing the mean-field Hamiltonian we find that a $d$-wave gap opens at the chemical potential. The resulting band structures for $\alpha = 0$ 
and ${pp}$ are shown in Fig.~\ref{gap} for a $100\times 100$ lattice at a density of 4.9 electrons per unit cell (5 electrons per unit cell corresponds to 
the undoped case) along the main directions in momentum space for various values of $\Delta$. Results for $\Delta=0$ are shown to indicate 
the non-interacting Fermi surface. The results for the on-site 
pairing operator $D_0$ ($\alpha = 0$) are shown in panel (a) 
of the figure, while those for the extended operator $D_{pp}$ ($\alpha = pp$) 
are in panel (b). In panels (c) and (d) it can be seen that 
for both $\Delta=0.3$ and 0.5 a gap opens at the antinodal position $X$ but the node along the 
diagonal direction $\Gamma-M$ remains, indicating the $d$-wave symmetry of the gap, as expected. 
In addition, note that the interaction only 
distorts the bands close to the Fermi surface and we observe a very flat dispersion of the band that defines the gap at $X$, in agreement with
recent experiments~\cite{smallpairs}. The results for the plaquette, inter-orbital, and traditional operators look very similar and are not shown explicitly.

\subsection{Stability of $d$-wave state}\label{estab}

The next aspect to explore is the stability of the pairing state with a finite gap. 
To study this issue we need to evaluate the energy of the mean-field Hamiltonian vs $\Delta$ 
for different values of the pairing strength $V$,
where  $\Delta=V\langle p_{-{\bf k},\mu,\downarrow}p_{{\bf k},\mu,\uparrow}\rangle$ for the on-site pairing $D_0$,
which we assume is the same for all values of $\mu$. The total energy is
\begin{equation}
E=\sum_{\bf k}[\sum_{i=1}^3(\epsilon_i({\bf k})-\mu_e)-E_i({\bf k})]+{\Delta^2N\over{V}},
\label{energy}
\end{equation}
\noindent where $\epsilon_i({\bf k})$ are the eigenvalues of the tight-binding term, $E_i$ are the three negative eigenvalues of the 
mean-field matrix (where the chemical potential has been included), and $N$ is the number of sites 
of the large but finite cluster used. The appropriate fermionic operators need to be used in the expression of $\Delta$ for the remaining $d$-wave pairing operators.

\begin{figure}[thbp]
\begin{center}
\includegraphics[width=8.5cm,clip,angle=0]{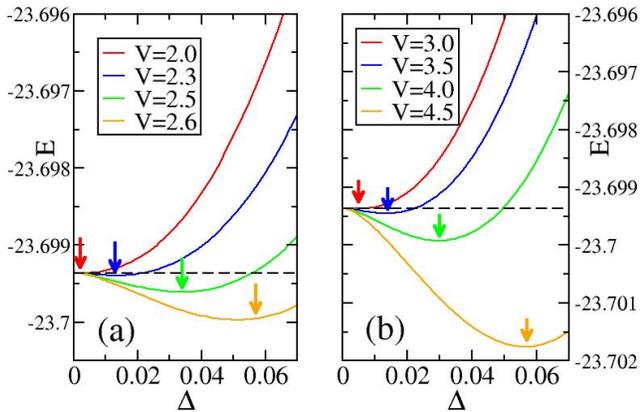}
\vskip -0.3cm
\caption{(color online) Total energy versus $\Delta$ at various values of $V$ for 
(a) the on-site $D_0$ and 
(b) the extended $D_{pp}$ pairing operators. 
Arrows indicate the minima in the energies. 
$\Delta$ and $E$ are in units of eV.} 
\vskip -0.4cm
\label{evsgap}
\end{center}
\end{figure}

We have observed that any finite value of $V$ stabilizes the proposed pairing states, similar to what happens in the negative-$U$ Hubbard model.
The small values of $\Delta$ that minimizes the energy for the different values of $V$ 
are indicated with an arrow in panels (a) and (b) of Fig.~\ref{evsgap} for the pairing operators $D_0$ and $D_{pp}$. 
Experimentally, the value of the superconducting 
gap in the cuprates ranges from 20 meV to 40 meV~\cite{shen}. Since the gap in our model is equal to $2\Delta$ 
we see from the figure that $V\sim 2.4$ ($V\sim 3.6$) provides a 
reasonable value of $\Delta$ for the minimum energy for on-site (extended) pairing.     
While all the $d$-wave pairing operators open a gap in the density of states as soon as $V$ is finite, we observed that all the pairing states that include $d$-orbitals 
produce larger gaps than the pure $p$-orbital operators at a fixed value of the attraction $V$. This becomes clear as we obtain the value of $V$ needed in each case to
open a gap similar to the one observed in the cuprates.  
In Table~\ref{table} we present the values of the attraction that stabilizes a gap of 
20~meV in each case, and it can be seen that $V<1$ eV ($V>1$ eV) is needed for operators that (do not) involve 
$d$-orbitals. 

\begin{table}
\begin{tabular}{|c|c|}\hline
Operator label & $V_{\Delta=20~{\rm meV}}$(eV)\\
\hline
$D_0$  & 2.38 \\
$D_{pp}$  & 3.68 \\
$D_{plaq}$  &3.50 \\
$D_{pd}$ &  0.50\\
$D_D$ & 0.13\\
$D_{D3B}$ & 0.15\\ \hline
 \end{tabular}
\caption{$V_{\Delta=20~{\rm meV}}$ indicates the value of the attraction that produces a total gap of 40 meV for the corresponding $d$-wave pairing.}

 \label{table}
 \end{table}

Thus, the mean-field results appear to indicate that pairing operators that involve the $d$-orbital need a much smaller attraction to produce a superconducting gap similar to the one observed in the cuprates. This is probably due 
to the fact that in the mean-field calculations the gap opens around the non-interacting Fermi surface which, as shown in Fig.~\ref{TBdisp}, it is mostly a $d$-band. In the cuprates though, it is expected that the
band that forms the non-interacting Fermi surface would generate upper and lower bands, due to the Coulomb repulsion at the Cu, and the Fermi surface upon doping will occur in a $p-d$ hybrizided band, identified as a 
Zhang-Rice band in photoemission~\cite{zaanen}. Due to the higher weight of the $p$-orbitals in this band, it is expected that the $p$-based compact $d$-wave order parameters proposed here may work better than the traditionally used pairing 
operators.

\section{Phenomenological Model} \label{effective}

Finally, if we replace $\Delta$ by $2V \gamma_{\nu}p_{{\bf j}+\hat\nu/2 ,\nu,\downarrow}p_{{\bf j}+\hat\nu/2,\nu,\uparrow}$ instead of the average value of the pairing operator in Eq.~(\ref{hintreals}), we obtain a phenomenological
interaction that should promote the on-site $d$-wave pairing $D_0$ given by:
\begin{equation}
\begin{split}
H_{\rm int}=\\-4V\sum_{{\bf j},\mu,\nu}\gamma_{\mu}\gamma_{\nu}p^{\dagger}_{{\bf j}+\hat\mu/2,\mu,\uparrow}p^{\dagger}_{{\bf j}+\hat\mu/2,\mu,\downarrow}p_{{\bf j}+\hat\nu/2 ,\nu,\downarrow}p_{{\bf j}+\hat\nu/2,\nu,\uparrow}=\\
-4V\sum_{{\bf j},\mu}n_{{\bf j}+\hat\mu/2,\mu,\uparrow}n_{{\bf j}+\hat\mu/2,\mu,\downarrow}+\\
4V\sum_{{\bf j},\mu\ne\nu}p^{\dagger}_{{\bf j}+\hat\mu/2,\mu,\uparrow}p_{{\bf j}+\hat\nu/2 ,\nu,\downarrow}
p^{\dagger}_{{\bf j}+\hat\mu/2,\mu,\downarrow}p_{{\bf j}+\hat\nu/2,\nu,\uparrow}.
\label{coulomb}
\end{split}
\end{equation}
The first term is an effective on-site attraction in the O sites while the second 
term involves the four O's that surround the Cu at site ${\bf j}$ and is repulsive.
While it is unlikely that terms of this form could be dynamically generated by the 
long-range Coulomb repulsion and a short-range attraction induced by antiferromagnetic fluctuations, it is important to remember that the electron-phonon interaction in
BCS superconductors does not lead to the instantaneous on-site attraction of the negative-$U$ Hubbard model. However, this model has been an important phenomenological tool to study
the behavior of $s$-wave superconductors, both with weak and strong attraction. Then, it is possible that the Hamiltonian here proposed could play a similar role but for $d$-wave superconductors.
In order to observe long-range order with this very local pairing operator it may be necessary to use a considerably large value of $V$ to allow for a higher $p-d$ hybridization at the Fermi surface.

\section{Conclusions}\label{conclu}

Summarizing, in this effort new intracell-CuO$_2$ -- three intra-orbital and one inter-orbital -- pairing operators 
with $d$-wave symmetry have been proposed for the high critical temperature cuprates. 
These operators are more local (more compact in size) 
than those previously employed in numerical studies and, thus, they may produce a stronger 
signal when their long-range behavior in finite systems is studied with numerical many-body techniques. 
In addition, they may be able to account for the small size of the pairs recently experimentally observed in the cuprates via
angle-resolved photoemission methods. At the 
mean-field level, the flatness of the band that forms the gap at the antinodes is reproduced 
and it is demonstrated that the size of the superconducting gap 
experimentally observed is obtained even with a moderate attraction~\cite{foot2}. 
The next step would be to evaluate more properties of these pairing operators at 
the mean-field level and, even more importantly, to calculate their pairing 
correlations in the three-orbital 
Hubbard model employing unbiased computational techniques.
    
\section{Acknowledgments}

The authors were supported by the U.S. Department of Energy (DOE), 
Office of Science, Basic Energy Sciences (BES), Materials
Sciences and Engineering Division.
\appendix
\section{$\lambda_i$ matrices}
\label{app:1_eff}

The $\lambda_i$ matrices used in the text are presented here:

\begin{align*}
  \lambda_0=
  \left(\begin{array}{ccc}
      1 & 0               &   0\\
      0 & 1               &   0\\
      0 & 0               &   1
    \end{array} \right),
  &\quad
  \lambda_1=
  \left(\begin{array}{ccc}
      0 & 1               &   0\\
      1 & 0               &   0\\
      0 & 0               &   0
    \end{array} \right),
\end{align*}

\begin{align*}
  \lambda_2=
  \left(\begin{array}{ccc}
      0 & -i               &   0\\
      i & 0               &   0\\
      0 & 0               &   0
    \end{array} \right),&\quad
  \lambda_3=
  \left(\begin{array}{ccc}
      1 & 0               &   0\\
      0 & -1               &   0\\
      0 & 0               &   0
    \end{array} \right),
\end{align*}

\begin{align*}
  \lambda_4=
  \left(\begin{array}{ccc}
      0 & 0               &   1\\
      0 & 0               &   0\\
      1 & 0               &   0
    \end{array} \right),&\quad
  \lambda_5=
  \left(\begin{array}{ccc}
      0 & 0               &   -i\\
      0 & 0               &   0\\
      i & 0               &   0
    \end{array} \right),
\end{align*}

\begin{align*}
  \lambda_6=
  \left(\begin{array}{ccc}
      0 & 0               &   0\\
      0 & 0               &   1\\
      0 & 1               &   0
    \end{array} \right),&\quad
  \lambda_7=
  \left(\begin{array}{ccc}
      0 & 0               &   0\\
      0 & 0               &   -i\\
      0 & i               &   0
    \end{array} \right),
\end{align*}

\begin{equation*}
  \lambda_8=\frac{1}{\sqrt{3}}
  \left(\begin{array}{ccc}
      1 & 0               &   0\\
      0 & 1               &   0\\
      0 & 0               &   -2
    \end{array} \right).
\end{equation*}


\begin{thebibliography}{}
\bibitem{foot2} While in BCS superconductors an infinitesimal attraction stabilizes the superconducting state, 
this may not be the case in the cuprates and, in addition, when the Coulomb interaction is taken into account
the orbital content at the chemical potential for a doped system will contain more $p$ contributions than in the non-interacting case.
\end{thebibliography}
\end{document}